\PassOptionsToPackage{unicode}{hyperref}
\PassOptionsToPackage{hyphens}{url}
\PassOptionsToPackage{dvipsnames,svgnames,x11names}{xcolor}
\documentclass[
]{article}
\usepackage{xcolor}
\usepackage[a4paper,margin=1in]{geometry}
\usepackage{amsmath,amssymb}
\usepackage{iftex}
\ifPDFTeX
  \usepackage[T1]{fontenc}
  \usepackage[utf8]{inputenc}
  \usepackage{textcomp} 
\else 
  \usepackage{unicode-math} 
  \defaultfontfeatures{Scale=MatchLowercase}
  \defaultfontfeatures[\rmfamily]{Ligatures=TeX,Scale=1}
\fi
\usepackage{lmodern}
\ifPDFTeX\else
\fi
\IfFileExists{upquote.sty}{\usepackage{upquote}}{}
\IfFileExists{microtype.sty}{
  \usepackage[]{microtype}
  \UseMicrotypeSet[protrusion]{basicmath} 
}{}
\makeatletter
\@ifundefined{KOMAClassName}{
  \IfFileExists{parskip.sty}{%
    \usepackage{parskip}
  }{
    \setlength{\parindent}{0pt}
    \setlength{\parskip}{6pt plus 2pt minus 1pt}}
}{
  \KOMAoptions{parskip=half}}
\makeatother
\usepackage{longtable,booktabs,array,tabularx,pdflscape,ragged2e}
\usepackage{authblk}
\usepackage{calc} 
\usepackage{etoolbox}
\makeatletter
\patchcmd\longtable{\par}{\if@noskipsec\mbox{}\fi\par}{}{}
\makeatother
\IfFileExists{footnotehyper.sty}{\usepackage{footnotehyper}}{\usepackage{footnote}}
\makesavenoteenv{longtable}
\usepackage{graphicx}
\makeatletter
\newsavebox\pandoc@box
\newcommand*\pandocbounded[1]{
  \sbox\pandoc@box{#1}%
  \Gscale@div\@tempa{\textheight}{\dimexpr\ht\pandoc@box+\dp\pandoc@box\relax}%
  \Gscale@div\@tempb{\linewidth}{\wd\pandoc@box}%
  \ifdim\@tempb\p@<\@tempa\p@\let\@tempa\@tempb\fi
  \ifdim\@tempa\p@<\p@\scalebox{\@tempa}{\usebox\pandoc@box}%
  \else\usebox{\pandoc@box}%
  \fi%
}
\def\fps@figure{htbp}
\makeatother
\NewDocumentCommand\citeproctext{}{}
\NewDocumentCommand\citeproc{mm}{%
  \begingroup\def\citeproctext{#2}\cite{#1}\endgroup}
\makeatletter
 \let\@cite@ofmt\@firstofone
 \def\@biblabel#1{}
 \def\@cite#1#2{{#1\if@tempswa , #2\fi}}
\makeatother
\newlength{\cslhangindent}
\newlength{\csllabelwidth}
\newenvironment{CSLReferences}[2] 
 {\begin{list}{}{%
  \setlength{\itemindent}{0pt}
  \setlength{\leftmargin}{0pt}
  \setlength{\parsep}{0pt}
  \ifodd #1
   \setlength{\leftmargin}{\cslhangindent}
   \setlength{\itemindent}{-1\cslhangindent}
  \fi
  \setlength{\itemsep}{#2\baselineskip}}}
 {\end{list}}
\usepackage{calc}

\providecommand{\tightlist}{%
  \setlength{\itemsep}{0pt}\setlength{\parskip}{0pt}}
\usepackage{bookmark}
\IfFileExists{xurl.sty}{\usepackage{xurl}}{} 
\hypersetup{
  colorlinks=true,
  linkcolor={black},
  filecolor={Maroon},
  citecolor={MidnightBlue},
  urlcolor={NavyBlue},
  pdfcreator={LaTeX via pandoc}}

\newcolumntype{Y}{>{\RaggedRight\arraybackslash}X}
\newcommand{\metric}[1]{\emph{#1}}
\newcommand{\businessutility}{\emph{Business utility}}

\title{Business Utility of Large Language Models as Exploratory Data Analysis Agents}
\author[1,2]{Rafał Łabędzki}
\author[1,3]{Patryk Miziuła}
\author[1]{Hubert Rutkowski}
\author[1]{Szymon Betlewski}
\author[1]{Cezary Depta}
\author[1]{Szymon Janowski}
\author[4]{Jarosław Kochanowicz}
\author[4]{Jan Kanty Milczek}
\affil[1]{deepsense.ai}
\affil[2]{SGH Warsaw School of Economics}
\affil[3]{Bydgoszcz University of Science and Technology}
\affil[4]{At the time of contribution affiliated with deepsense.ai, present affiliation is Google}
\date{May 2026}

\begin{document}
\maketitle

\begin{abstract}

Large Language Models (LLMs) are increasingly used in analytical
workflows, but their suitability as exploratory data analysis (EDA)
agents in business settings remains uncertain. In practice, a deployable
EDA agent must provide not only useful average performance but also
sufficient repeatability to support trust in its outputs. We evaluate
this requirement in a controlled, business-relevant benchmark built on
an agent-based supply chain simulation. The task is to identify
supplier-product combinations responsible for low quality and downstream
sales loss by reasoning from indirect operational traces rather than
from explicit labels. Fifteen model-variant configurations from eight
model families were evaluated under four experimental conditions that
varied data representation, prompt clarity, and signal strength, with
five trajectories per condition. Outputs were scored against
deterministic ground truth using the Jaccard index and assessed through
a framework that combines mean score (\metric{ms}), coefficient of
variation (\metric{CV}), exploratory cross-condition significance tests,
and \businessutility{}, a risk-adjusted metric that we propose
to summarize quality and repeatability in a single operational measure.
The results show that most configurations are not reliable enough for
autonomous EDA use, even when their average scores appear acceptable.
GPT-5.4 with extra-high reasoning effort achieved the strongest overall
profile, with an experiment-averaged \metric{ms} of 0.8748 and an
experiment-averaged \businessutility{} of 0.6952, while the next-best
configurations lost substantially
more utility after variability discounting. Our findings suggest that
evaluation of EDA agents should treat average quality, repeatability,
and condition sensitivity as complementary dimensions of operational
trustworthiness.

\end{abstract}

\section{Introduction}\label{introduction}

Large Language Models have become central components of contemporary AI
systems. They are increasingly embedded in agentic workflows,
retrieval-augmented pipelines, and interactive analytical tools that
allow them to act on external environments rather than merely respond in
natural language (\citeproc{ref-gao_retrieval-augmented_2024}{Gao et al.
2024}; \citeproc{ref-li_survey_2024}{Li et al. 2024}). This shift has
intensified interest in whether LLMs can assist or automate specialist
tasks that require structured reasoning, procedural execution, and
interpretation.

Data analysis is a particularly important case. Existing work shows that
LLMs can assist with code generation, notebook completion, and selected
data science subtasks, but strong evidence for their use in exploratory
data analysis remains limited and fragmented
(\citeproc{ref-jansen_leveraging_2025}{Jansen et al. 2025};
\citeproc{ref-lai_ds-1000_2022}{Lai et al. 2022}). That gap matters
because EDA is not simply a question-answering exercise. It is an
iterative analytical process in which the analyst describes data,
identifies salient structure, and interprets that structure in relation
to the underlying problem (\citeproc{ref-de_mast_exploratory_2007}{De
Mast and Trip 2007}; \citeproc{ref-tukey_exploratory_1977}{Tukey 1977}).

This research asks whether an LLM embedded in an agentic harness can
function as a dependable EDA agent under realistic business conditions.
In organizational settings, a model is useful only if it combines acceptable average
performance with low variability, because high volatility increases
verification costs and makes failures difficult to anticipate. For that
reason, our evaluation extends beyond benchmark means to include
repeatability and sensitivity to realistic changes in working
conditions. To formalize that perspective, the paper introduces a
synthetic metric, \businessutility{}, that aggregates
performance and reliability in a single value between \texttt{0} and
\texttt{1}. The metric treats mean score as expected analytical value
and discounts that value by instability through a
prospect-theory-inspired loss function over variability, with that
relation discussed later in the evaluation framework
(\citeproc{ref-kahneman_tversky_1979}{Kahneman and Tversky 1979};
\citeproc{ref-tversky_kahneman_1992}{Tversky and Kahneman 1992}).

This paper presents a benchmark built around a realistic business
scenario. We evaluate 15 model-variant configurations on an agent-based
simulation of a supply chain in which the task is to infer which
supplier-product combinations caused low quality and downstream sales
loss. The paper makes three contributions. First, it introduces a
controlled benchmark that operationalizes EDA as a causal reconstruction
problem over business-process data. Second, it proposes a risk-adjusted
evaluation framework that combines mean task quality with variability.
Third, it provides empirical evidence that current LLMs remain
operationally fragile as autonomous EDA agents, with instability
emerging as the main obstacle to dependable deployment.

\section{Background: Exploratory Data
Analysis}\label{background-exploratory-data-analysis}

Exploratory data analysis is commonly understood as a process of
learning from data before confirmatory testing. Tukey
(\citeproc{ref-tukey_exploratory_1977}{1977}) framed EDA as a way
to let data suggest patterns, anomalies, and hypotheses rather than
merely confirm prior expectations. Later work emphasized that EDA is not
reducible to descriptive statistics alone. Instead, it combines data
display, pattern discovery, anomaly detection, and interpretation in a
form that aligns analytical representations with human cognitive
strengths (\citeproc{ref-good_philosophy_1983}{Good 1983};
\citeproc{ref-morgenthaler_exploratory_2009}{Morgenthaler 2009}).

For the present study, the most useful formulation is the three-step
view proposed by De Mast and Trip
(\citeproc{ref-de_mast_exploratory_2007}{2007}):
display the data, identify salient features, and interpret salient
features. We adopt a slightly expanded wording in which the first step
is expressed as \emph{describe the data}. This does not separate
descriptive data analysis from EDA as a fully independent phase. Rather,
it makes explicit that exploratory work begins by organizing,
summarizing, and displaying data in a way that allows salient variation
to become visible (\citeproc{ref-allen_exploratory_2018}{Allen et al.
2018}; \citeproc{ref-li_vigni_exploratory_2013}{Li Vigni et al. 2013}).
Under this interpretation, EDA can be described as a three-part process:

\begin{enumerate}
\def\labelenumi{\arabic{enumi}.}
\tightlist
\item
  describe the data,
\item
  identify salient features,
\item
  interpret salient features.
\end{enumerate}

This process orientation matters because the endpoint of EDA is not
merely a table, a visualization, or a classification label. The endpoint
is a set of plausible explanatory hypotheses about which variables or
mechanisms deserve further attention. De Mast and Trip
(\citeproc{ref-de_mast_exploratory_2007}{2007}) treat
this as a hypothesis-generation problem in which patterns, deviations,
or structural irregularities in the data motivate subsequent
interpretation.

The benchmark in this paper is built around that conception. The solver
is not given a clean label indicating which producer is defective.
Instead, the solver must inspect operational records, reconstruct how
goods moved through the simulated system, detect the traces of quality
failure in downstream behavior, and infer which supplier-product pairs
plausibly generated those traces. The benchmark therefore treats EDA as
structured hypothesis generation rather than as isolated code generation
or closed-form question answering.

\section{Related Work}\label{related-work}

\subsection{Benchmarking Data and ML
Agents}\label{benchmarking-data-and-ml-agents}

The number of benchmarks for LLMs and agents in data science, analytics,
and machine learning engineering has grown rapidly in recent years.
However, these benchmarks differ substantially in what they test, how
they test it, and how close they come to process-oriented EDA. Some
focus on code generation or closed-form data-analysis tasks, while
others evaluate longer analytical workflows, insight discovery, or
tool-grounded agent behavior in realistic environments.

From the perspective of this paper, the most important distinction is
between benchmarks that are directly relevant to exploratory or
insight-oriented analysis and benchmarks that primarily target machine
learning engineering. That distinction matters because a benchmark can
be valuable while still leaving the core process of EDA under-specified.

Table 1 summarizes the benchmark landscape that informed this study.

Table 1. \textbf{Comparison of benchmarks evaluating ML and data-science
agent capabilities}

\begin{longtable}[]{@{}
  >{\raggedright\arraybackslash}p{(\linewidth - 10\tabcolsep) * \real{0.1667}}
  >{\raggedright\arraybackslash}p{(\linewidth - 10\tabcolsep) * \real{0.1667}}
  >{\raggedright\arraybackslash}p{(\linewidth - 10\tabcolsep) * \real{0.1667}}
  >{\raggedright\arraybackslash}p{(\linewidth - 10\tabcolsep) * \real{0.1667}}
  >{\raggedright\arraybackslash}p{(\linewidth - 10\tabcolsep) * \real{0.1667}}
  >{\raggedright\arraybackslash}p{(\linewidth - 10\tabcolsep) * \real{0.1667}}@{}}
\toprule\noalign{}
\begin{minipage}[b]{\linewidth}\raggedright
Benchmark
\end{minipage} & \begin{minipage}[b]{\linewidth}\raggedright
Release
\end{minipage} & \begin{minipage}[b]{\linewidth}\raggedright
Benchmark goal
\end{minipage} & \begin{minipage}[b]{\linewidth}\raggedright
Evaluation approach
\end{minipage} & \begin{minipage}[b]{\linewidth}\raggedright
Number of problems
\end{minipage} & \begin{minipage}[b]{\linewidth}\raggedright
Origin of problems
\end{minipage} \\
\midrule\noalign{}
\endhead
\bottomrule\noalign{}
\endlastfoot
\textbf{2022} & & & & & \\
DS-1000 (\citeproc{ref-lai_ds-1000_2022}{Lai et al. 2022}) & Nov 2022 &
Data science & Execution semantics and test-case constraints determine
task success & 1000 & Public domain with modifications \\
\textbf{2024} & & & & & \\
InfiAgent-DABench (\citeproc{ref-hu_infiagent-dabench_2024}{Hu et al.
2024}) & Jan 2024 & Data science & Closed-form automatic scoring through
format prompting and code execution & 311 & Public datasets and genuine
problems not available in the public domain \\
BIBench (\citeproc{ref-liu_bibench_2024}{Liu et al. 2024}) & Feb 2024 &
Data science, business intelligence & Classification, extraction, and
generation tasks across knowledge, application, and analytical skill &
11 & Public domain \\
DSEval (\citeproc{ref-zhang_benchmarking_2024}{Zhang et al. 2024}) & Feb
2024 & Data science & Bootstrapped annotations, validator feedback on
code execution, and oracle-agent comparison & 4 benchmarks & Public
domain \\
MLAgentBench (\citeproc{ref-huang_mlagentbench_2024}{Q. Huang et al.
2024}) & Jun 2024 & Machine learning engineering & Success rate,
efficiency, and evaluator scoring in end-to-end experimentation tasks &
13 & Public domain \\
Spider2-V (\citeproc{ref-cao_spider2-v_2024}{Cao et al. 2024}) & Jul
2024 & Multimodal data-science and engineering workflows & Task-specific
programmatic verifiers in real desktop environments & 494 & Genuine
workflow tasks from tutorials and real projects \\
ML-Bench (\citeproc{ref-tang_ml-bench_2024}{Tang et al. 2024}) & Aug
2024 & Machine learning engineering & Pass@K for text-to-code and
success rate for end-to-end agent execution & 9641 & Public domain \\
DA-Code (\citeproc{ref-huang_da-code_2024}{Y. Huang et al. 2024}) & Oct
2024 & Data science & Execution-based correctness in a sandboxed
environment & 500 & Public datasets and genuine problems not available
in the public domain \\
SWE-bench (\citeproc{ref-jimenez_swe-bench_2024}{Jimenez et al. 2024}) &
Nov 2024 & Software engineering & Patch resolution and test-passing
outcomes on real issues & 2294 & Public domain \\
RE-Bench (Wijk et al., n.d.) & 2024 & Machine learning engineering &
Task-specific scores in open-ended environments normalized to human
baselines & 7 & Genuine datasets and problems \\
\textbf{2025} & & & & & \\
DSBench (\citeproc{ref-jing_dsbench_2025}{Jing et al. 2025}) & Feb 2025
& Data science & Semantic similarity, relative performance gap, and task
success rate & 540 & Public domain \\
DataSciBench (\citeproc{ref-zhang_datascibench_2025}{Zhang et al. 2025})
& Feb 2025 & Data science & Semi-automated LLM assessment combined with
human judgments and programmed rules & 222 & Public domain, human made,
and LLM generated \\
MLE-bench (\citeproc{ref-chan_mle-bench_2025}{Chan et al. 2025}) & Feb
2025 & Machine learning engineering & Offline Kaggle competition
performance compared with human submissions & 75 & Public domain \\
IDA-Bench (\citeproc{ref-li_ida-bench_2025}{Li et al. 2025}) & Jun 2025
& Interactive guided data analysis & Final performance against human
notebook baselines in multi-round workflows & 25 & Recent Kaggle
notebooks with curated task distillation \\
DABstep (\citeproc{ref-egg_dabstep_2025}{Egg et al. 2025}) & Jun 2025 &
Multi-step data analysis reasoning & Objective factoid scoring with
numeric tolerance and normalized string or list matching & 450+ &
Genuine anonymized industry analytical workloads \\
Text2Vis (\citeproc{ref-rahman_text2vis_2025}{Rahman et al. 2025}) & Jul
2025 & Text-to-visualization generation & Combined short-answer
correctness, code execution, and chart-quality criteria & 1985 & Curated
public tables and generated queries with human verification \\
NL2SQL-BUGs (\citeproc{ref-liu_nl2sql-bugs_2025}{Liu et al. 2025}) & Aug
2025 & NL2SQL semantic error detection & Binary and type-specific
semantic error detection & 2018 & Public-domain construction from BIRD
with expert annotations \\
FDABench (\citeproc{ref-wang_fdabench_2025}{Wang et al. 2025}) & Sep
2025 & Data analysis over heterogeneous data & Exact match, ROUGE, and
efficiency metrics across workflow types & 2007 & Public benchmarks
extended with curated unstructured context \\
InsightEval (\citeproc{ref-zhu_insighteval_2025}{Zhu et al. 2025}) & Nov
2025 & Insight discovery in EDA & Insight recall, precision, F1, and
novelty scoring & 100 instances & Expert-curated and human-verified
benchmark instances \\
DAComp (\citeproc{ref-lei_dacomp_2025}{Lei et al. 2025}) & Dec 2025 &
Full data-intelligence lifecycle & Execution-based scoring for
deterministic tasks and rubric-guided LLM judging for open-ended
analysis & 210 & Genuine enterprise-inspired schemas and curated
analytical databases \\
TimeSeriesGym (\citeproc{ref-cai_timeseriesgym_2025}{Cai et al. 2025}) &
2025 & Time-series ML engineering & Quantitative artifact metrics,
programmatic checks, and LLM-judge scoring & 34 & Public competitions
and curated repository tasks \\
nvBench 2.0 (\citeproc{ref-luo_nvbench_2025}{Luo et al. 2025}) & 2025 &
Ambiguous text-to-visualization reasoning & Precision@K, Recall@K, and
F1@K over multiple valid visualizations & 7878 & Public tables
transformed through controlled ambiguity injection \\
\textbf{2026} & & & & & \\
ML-Tool-Bench (\citeproc{ref-chittepu_ml-tool-bench_2026}{Chittepu et
al. 2026}) & Feb 2026 & Tool-augmented ML planning & Validity of
long-horizon tool trajectories and leaderboard percentile performance &
15 & Public Kaggle tasks with curation and subsampling \\
\end{longtable}

Several patterns are visible in this comparison. First, benchmark scale
varies dramatically. Some benchmarks contain only a few highly complex
open-ended tasks, whereas others contain thousands of more standardized
problems. This difference matters because a large benchmark is not
necessarily closer to process-oriented EDA than a smaller but more
realistic workflow benchmark. Second, task provenance varies from public
repositories, StackOverflow, and Kaggle competitions to expert-curated
industry-style problems. Publicly available tasks are useful for scale
and reproducibility, but they also carry a greater risk that similar
tasks may appear in model training data. Third, evaluation logic varies
from exact execution-based scoring to semantic similarity, human or
oracle baselines, insight metrics, and LLM-as-judge protocols.

\subsection{The Gap Addressed by This
Study}\label{the-gap-addressed-by-this-study}

Despite this growth, none of the reviewed benchmarks are dedicated
solely to full process-oriented EDA in the sense adopted in this paper.
Some benchmarks are clearly EDA-adjacent. \texttt{InsightEval} focuses
on insight discovery, \texttt{IDA-Bench} focuses on interactive guided
analysis, \texttt{DABstep} evaluates multi-step analytical reasoning,
and \texttt{DAComp} includes open-ended data-analysis tasks. Although these are
important advances, they do not fully cover the specific EDA
sequence emphasized by De Mast and Trip
(\citeproc{ref-de_mast_exploratory_2007}{2007}):
describing data, identifying salient features, and interpreting them in
relation to an explanatory problem.

This gap motivates the present benchmark. The task is not primarily
about producing code, generating a chart, or passing a closed-form
verifier. It is about reconstructing a hidden causal pattern from
business-process data. That makes the study relevant to the broader
literature on data science agents while also targeting a narrower
problem, whether LLMs can function as dependable exploratory data
analysts when success requires both inference quality and repeatability
to build trust among decision makers and therefore increase a
probability of successful adoption in the organization.

\section{Methods}\label{methods}

\subsection{Research Objective}\label{research-objective}

The primary objective of the study is to assess the
\businessutility{} of LLM-based EDA agents in a controlled but
business-relevant analytical task. More specifically, the study asks
which model-variants combine useful task performance with enough
reliability to support trust in their outputs.

This objective is evaluated through a hierarchy of complementary
metrics:

\begin{enumerate}
\def\labelenumi{\arabic{enumi}.}
\tightlist
\item
  \metric{ms} (\texttt{mean\ score}) captures average analytical
  quality.
\item
  \metric{CV} (\texttt{Coefficient\ of\ Variation}) captures
  unpredictability.
\item
  \businessutility{} provides a risk-adjusted synthesis of
  quality and variability.
\item
  Condition sensitivity, operationalized through Mann-Whitney U and
  Kruskal-Wallis tests, indicates whether model behavior shifts across
  experimental conditions.
\end{enumerate}

Cross-experiment testing in this research is interpreted as an auxiliary
descriptor of model behavior alongside \metric{ms}, \metric{CV}, and
\businessutility{}.

\subsection{Benchmark Task and Simulation
Context}\label{benchmark-task-and-simulation-context}

The benchmark is built on an agent-based simulation of a supply chain
for everyday consumer goods. Products flow from producers through a
wholesaler to stores and then to customers. The practical problem arises
when some producers deliver low-quality batches. Poor quality does not
appear as a direct label in the data. Instead, it generates downstream
traces in customer demand, store orders, wholesaler allocations, and
store sales.

The solver is therefore required to analyze the available records,
reconstruct where goods reaching stores came from, connect supplier
deliveries with later customer behavior, and identify the producers who
were the real source of quality problems for particular goods. This
design makes the task analytically meaningful because it resembles a
realistic business diagnosis problem in which the cause of performance
degradation must be inferred from indirect operational evidence rather
than read off a single field.

\subsection{Experimental Design}\label{experimental-design}

All experiments used a supply chain simulation, but the analytical
environment was varied in ways intended to approximate realistic
differences in how a human analyst may encounter the same business
problem. The four experiments (we refer to experiments also as
conditions) were defined as shown in Table 2.

Table 2. \textbf{Experimental conditions}

\begin{longtable}[]{@{}
  >{\raggedright\arraybackslash}p{(\linewidth - 6\tabcolsep) * \real{0.2500}}
  >{\raggedright\arraybackslash}p{(\linewidth - 6\tabcolsep) * \real{0.2500}}
  >{\raggedright\arraybackslash}p{(\linewidth - 6\tabcolsep) * \real{0.2500}}
  >{\raggedright\arraybackslash}p{(\linewidth - 6\tabcolsep) * \real{0.2500}}@{}}
\toprule\noalign{}
\begin{minipage}[b]{\linewidth}\raggedright
Experiment
\end{minipage} & \begin{minipage}[b]{\linewidth}\raggedright
Data representation
\end{minipage} & \begin{minipage}[b]{\linewidth}\raggedright
Prompt
\end{minipage} & \begin{minipage}[b]{\linewidth}\raggedright
Simulation setup and signal
\end{minipage} \\
\midrule\noalign{}
\endhead
\bottomrule\noalign{}
\endlastfoot
\texttt{Experiment\ 1} (reference) & Raw data & Short & Simulation setup
1 with clear and strong signal \\
\texttt{Experiment\ 2} & Raw data and redundant tabular data & Short &
Simulation setup 1 with clear and strong signal \\
\texttt{Experiment\ 3} & Raw data & Non-ambiguous & Simulation setup 1
with clear and strong signal \\
\texttt{Experiment\ 4} & Raw data & Short & Simulation setup 2 with
clear but weaker signal \\
\end{longtable}

Each model-variant was evaluated on five trajectories per experiment,
for 20 trajectories in total. If a trajectory failed, one retry was
granted for that trajectory. If the retry also failed to produce an
output, the trajectory received a score of \texttt{0}. Failure was
defined as either missing JSON output or a stalled run with no
\texttt{stdout} or \texttt{stderr} change for 1200 seconds. All
trajectories were executed with the same OpenCode harness
(\texttt{v1.2.27}) and default setup, without additional custom
instructions such as \texttt{agents.md}. All models were evaluated with
their default temperature settings.

Most models were tested in low and high reasoning-effort variants
(called model-variants), while Mistral Large
exposed only a default configuration.

\subsection{Models and Output
Scoring}\label{models-and-output-scoring}

The benchmark covered eight LLM families and 15 model-variant
configurations. The output required from each run was a JSON object
containing the predicted set of supplier-good combinations associated
with the quality problem. Predictions were compared against
deterministic ground truth using the Jaccard score. This choice reflects
the practical objective of recovering all relevant supplier-good pairs
while avoiding false inclusions.

\subsection{Evaluation Framework}\label{evaluation-framework}

At the empirical level, each trajectory contributed two raw
measurements: task score and completion time. These values were then
aggregated by experiment and by model-variant. Three descriptors are
reported directly:

\begin{enumerate}
\def\labelenumi{\arabic{enumi}.}
\tightlist
\item
  \metric{ms}, which estimates average analytical correctness (Jaccard score).
\item
  \metric{CV}, which quantifies relative variation in score within the
  same condition.
\item
  \texttt{mean\ time}, which estimates average execution cost.
\end{enumerate}

At the interpretive level, these quantities are translated into a
risk-adjusted assessment. The motivation is operational rather than
purely descriptive. A model can have an acceptable mean score and still
be unsuitable for deployment if its behavior is too unstable. For that reason, the benchmark treats efficacy
and repeatability as jointly necessary conditions of usefulness for
business purposes.

The baseline quantity is \metric{ms}, the mean score observed for a
model in one experiment. This quantity captures expected analytical
usefulness, but it does not account for how reproducibly that usefulness
is delivered. To capture that second aspect, the benchmark uses the
coefficient of variation, $\mathrm{CV} = \sigma / \mathrm{ms}$. The
appeal of \metric{CV} is that it is scale-invariant, as it measures
relative rather than absolute instability, which makes comparisons more
meaningful across task settings with different baseline difficulty.

The study introduces the following metric:

\begin{equation}
\textit{Business utility} =
\mathrm{ms}\, D(\mathrm{CV}) =
\mathrm{ms}\, e^{-L(\mathrm{CV})} =
\mathrm{ms}\, e^{-2.25\,\mathrm{CV}^{0.88}}
\end{equation}

where \metric{ms} is the mean score and \metric{CV} is the coefficient
of variation. The metric can be decomposed into two parts. First,
instability is mapped into a prospect-theory-inspired loss term,
${L(\mathrm{CV}) = 2.25\,\mathrm{CV}^{0.88}}$. Second, that loss is
translated into a bounded stability discount,
${D(\mathrm{CV}) = e^{-L(\mathrm{CV})}}$.
\businessutility{} is therefore the product of expected
analytical usefulness and a smooth discount for instability.

This construction is designed to preserve three properties that are
important in the present benchmark. First, utility remains bounded
between $0$ and \metric{ms}, so higher instability always lowers
the summary without generating logically inconsistent negative values.
Second, the discount is nonlinear, which allows small departures from
ideal repeatability to matter more than they would under a purely
proportional penalty. Third, the multiplicative form keeps the
interpretation operationally simple: a model receives high utility only
when it is both accurate on average and sufficiently repeatable.

From a decision-theoretic perspective, the key design choice is the
specific instability loss term used in the discount function.
Its parameters are taken from the loss-side parametrization of cumulative
prospect theory, where $2.25$ captures loss aversion and
$0.88$ captures the curvature of the value function
(\citeproc{ref-tversky_kahneman_1992}{Tversky and Kahneman 1992}). In
the present benchmark, those parameters are not used to model monetary
outcomes directly. Instead, they are repurposed to express the idea that
departures from ideal repeatability should reduce perceived usefulness
in a nonlinear way. Under this interpretation, low instability remains
highly valuable, whereas increases in \metric{CV} are translated into a
progressively accumulated loss of trust.

For fixed \metric{CV}, the metric remains linear in \metric{ms} - each
increment in mean task quality contributes proportionally to
\businessutility{}. The resulting construction is therefore
hybrid - the instability term is prospect-theory-inspired, whereas the
efficacy term remains proportional and directly interpretable. This
separation is deliberate. It preserves a simple reading of \metric{ms}
as expected analytical usefulness while allowing instability to discount
that usefulness through a nonlinear loss-sensitive transformation.

One technical convention requires caution. For groups with
$\mathrm{ms} = 0$, we must have $\sigma = 0$ as
well and the existing aggregation logic sets $\mathrm{CV} = 0.0$ to
avoid undefined division. Under this convention,
\businessutility{} remains $0$, which is intuitive for
fully collapsed zero-score states, but the derived stability discount is
not substantively informative in those cases.

Cross-condition significance testing serves a different purpose.
Two-sided Mann-Whitney U tests compare score distributions between the
reference condition and selected interventions, while the Kruskal-Wallis
test evaluates whether a model-variant exhibits any detectable
distribution shift across all four experiments. Because each comparison
is based on a very small sample and the study evaluates multiple
model-specific contrasts without multiplicity correction, the resulting
$p$ values are interpreted as exploratory diagnostic signals
rather than confirmatory evidence.

\section{Results}\label{results}

\subsection{Mean performance across
experiments}\label{mean-performance-across-experiments}

The benchmark covered 4 experiments, with 5 trajectories for each 15
model-variants (from 8 LLM families), producing 300 scored trajectories
in total, with 75 trajectories per experiment. The overall \metric{ms}
across all trajectories was \texttt{0.3562}, and mean runtime was
\texttt{783.56} seconds. Performance was highest in the reference
experiment and lowest in the weaker-signal condition.

Table 3. \textbf{Mean performance across experiments}

\begin{longtable}[]{@{}llll@{}}
\toprule\noalign{}
Scope & n trajectories & \metric{ms} & Mean time (s) \\
\midrule\noalign{}
\endhead
\bottomrule\noalign{}
\endlastfoot
Overall & 300 & 0.3562 & 783.56 \\
Experiment 1 & 75 & 0.4088 & 774.84 \\
Experiment 2 & 75 & 0.3972 & 650.54 \\
Experiment 3 & 75 & 0.3936 & 751.00 \\
Experiment 4 & 75 & 0.2253 & 957.86 \\
\end{longtable}

The strongest aggregate degradation appears in Experiment 4,
where the underlying signal is weaker and the simulation setup changes.
By contrast, the addition of redundant tabular aggregates in
Experiment 2 and the non-ambiguous prompt in
Experiment 3 do not materially shift the pooled \metric{ms}.

\subsection{Business utility based
ranking}\label{business-utility-based-ranking}

The ranking of all 15 evaluated model-variants by
experiment-averaged \businessutility{} is presented in Table 4. The impact of 
model variability on their position in the ranking is clearly visible. Some models with lower \metric{ms} 
above models with higher \metric{ms} because of their better stability. For example,
Gemini 3.1 Pro high remains above GPT-5.2 xhigh because its average \metric{CV}
is lower and therefore its average stability discount is weaker.

\begin{table}[!htbp]
\centering
\small
\caption{Global ranking by experiment-averaged \businessutility{}. Model names are shortened for display.}
\label{tab:business-utility-ranking}
\begin{tabular*}{\linewidth}{@{\extracolsep{\fill}}r l l r r r r@{}}
\toprule
Rank & Model & Variant & Avg.\ \metric{ms} & Avg.\ \metric{CV} & Avg.\ disc. & Avg.\ Business utility \\
\midrule
1 & GPT-5.4 & xhigh & 0.8748 & 0.1202 & 0.7665 & 0.6952 \\
2 & Gemini 3.1 Pro & high & 0.4991 & 0.2622 & 0.6031 & 0.3533 \\
3 & GPT-5.2 & xhigh & 0.6501 & 0.3739 & 0.4126 & 0.2815 \\
4 & Claude Sonnet 4.6 & low & 0.4597 & 0.3434 & 0.4728 & 0.2506 \\
5 & Claude Haiku 4.5 & high & 0.3850 & 0.4361 & 0.3573 & 0.1513 \\
6 & Claude Haiku 4.5 & low & 0.3620 & 0.6022 & 0.2586 & 0.1044 \\
7 & GPT-5.4 & low & 0.3730 & 0.6973 & 0.2431 & 0.1016 \\
8 & Claude Sonnet 4.6 & high & 0.3432 & 0.7240 & 0.2263 & 0.0966 \\
9 & GPT-5.4 mini & xhigh & 0.4252 & 0.8091 & 0.1934 & 0.0929 \\
10 & Gemini 3 Flash & high & 0.3419 & 0.7755 & 0.1799 & 0.0700 \\
11 & Gemini 3 Flash & low & 0.2501 & 1.1577 & 0.0926 & 0.0289 \\
12 & GPT-5.2 & low & 0.1128 & 0.6526 & 0.5622 & 0.0258 \\
13 & Mistral Large & default & 0.1635 & 1.1297 & 0.0839 & 0.0130 \\
14 & Gemini 3.1 Pro & low & 0.1003 & 2.0000 & 0.0159 & 0.0016 \\
15 & GPT-5.4 mini & low & 0.0027 & 0.5000 & 0.7540 & 0.0001 \\
\bottomrule
\end{tabular*}
\end{table}

The ranking places GPT-5.4 xhigh clearly above
the remaining configurations under the revised discount. Its
experiment-averaged \metric{ms} remains the highest, and its
experiment-averaged \metric{CV} of \texttt{0.1202} still preserves a
strong average stability discount of \texttt{0.7665}, leaving an
experiment-averaged \businessutility{} of \texttt{0.6952}. The
next tier consists of Gemini 3.1 Pro
high, GPT-5.2 xhigh, and
Claude Sonnet 4.6 low. The change in
ordering between the latter two is substantive - under the
prospect-theory-inspired discount, GPT-5.2
xhigh recovers enough utility from its stronger mean
performance to overtake the more stable but lower-scoring
Claude Sonnet 4.6 low.

\begin{figure}[!t]
\centering
\includegraphics[width=\linewidth,height=0.72\textheight,keepaspectratio,alt={Model-variant performance profiles across experiments}]{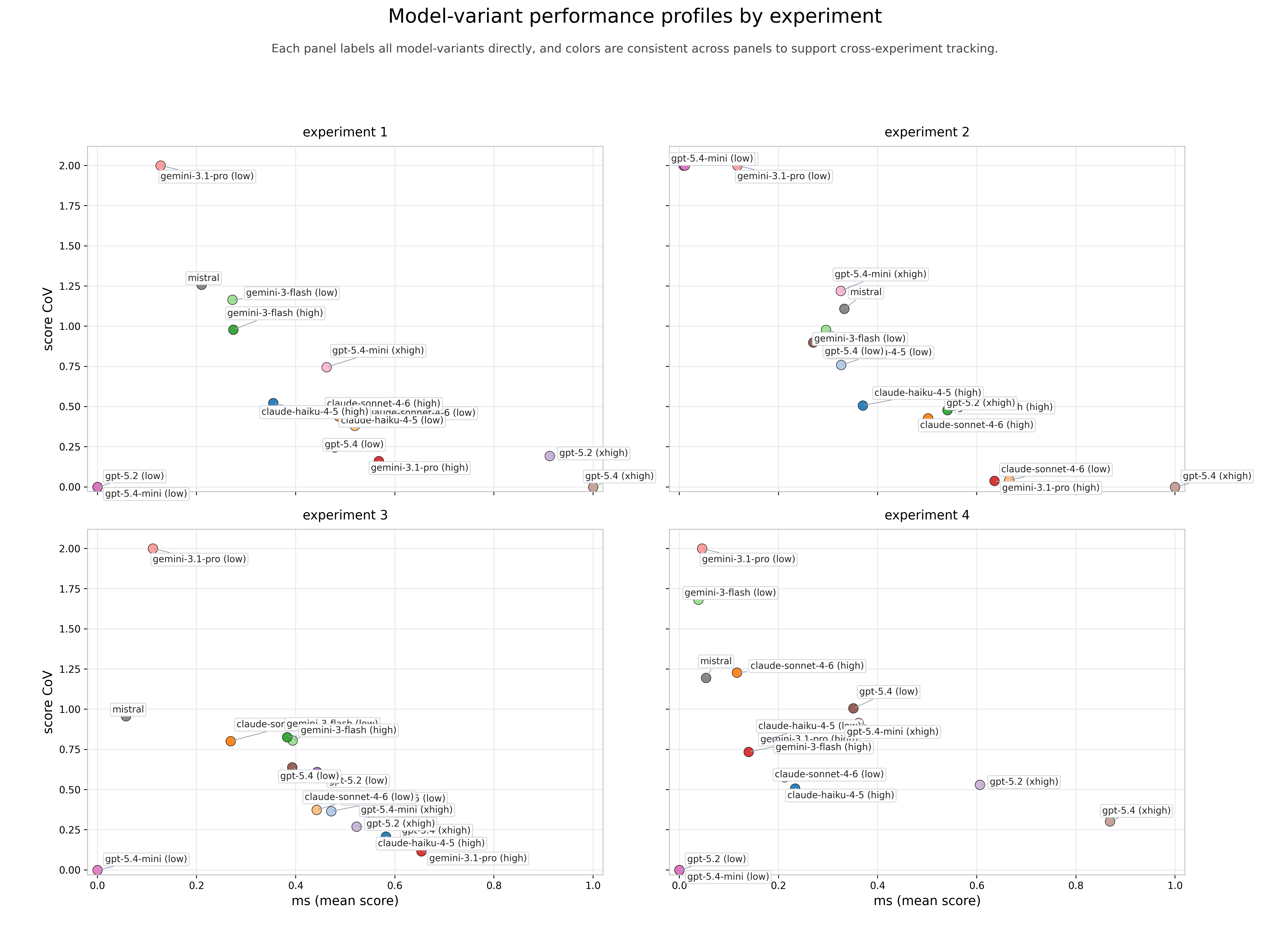}
\caption{Model-variant performance profiles across experiments}
\label{fig:model-variant-profiles}
\end{figure}

The full per-experiment profile is shown in Figure~\ref{fig:model-variant-profiles}. 
The chart reinforces three conclusions. First, only a small subset of
configurations occupies a clearly favorable region with both high
\metric{ms} and low \metric{CV}. Second, the relative positions of the
same variants differ across the four panels rather than following one
common directional pattern, which is consistent with the selective
condition-sensitivity results. Third, regardless the experiment, many 
model-variants occupy high-variability regions, confirming that
within-condition instability remains a central obstacle to reliable
deployment.

\subsection{Condition Sensitivity}\label{condition-sensitivity}

Condition sensitivity is visible but not strong enough to eclipse the
larger issue of within-condition variability. At the pooled level, only
the comparison between Experiment 1 and Experiment 4
crosses $p \le 0.05$ threshold.

\begin{table}[!htbp]
\centering
\small
\caption{Pooled condition sensitivity across experiments. Asterisks indicate $p \le 0.05$.}
\label{tab:pooled-condition-sensitivity}
\begin{tabular*}{\linewidth}{@{\extracolsep{\fill}}r r r r r r@{}}
\toprule
\metric{ms} & Mean time (s) & MWU 1 vs 2 & MWU 1 vs 3 & MWU 1 vs 4 & KW 1--4 \\
\midrule
0.3562 & 783.56 & 0.811397 & 0.908204 & 0.000908* & 0.000251* \\
\bottomrule
\end{tabular*}
\end{table}

At the model-variant level, condition sensitivity is selective rather
than universal. No model-variant crosses the
$p \le 0.05$ threshold for the comparison between
Experiment 1 and Experiment 2, which is consistent
with the view that adding redundant tabular aggregates did not
systematically change performance in this dataset. Three variants cross
the threshold for Experiment 3, and two variants cross it for
Experiment 4. Six variants also cross the threshold in the
omnibus Kruskal-Wallis screen across all four experiments.

\begin{landscape}
\begin{table}[p]
\centering
\small
\caption{Model-variant condition sensitivity. Asterisks indicate $p \le 0.05$. Model names are shortened for display.}
\label{tab:model-condition-sensitivity}
\begin{tabular*}{\linewidth}{@{\extracolsep{\fill}}l l r r r r r r@{}}
\toprule
Model & Variant & \metric{ms} & Mean time & MWU 1 vs 2 & MWU 1 vs 3 & MWU 1 vs 4 & KW 1--4 \\
\midrule
Claude Haiku 4.5 & high & 0.3850 & 393.53 & 1.000000 & 0.095238 & 0.397615 & 0.044408* \\
Claude Haiku 4.5 & low & 0.3620 & 321.77 & 0.396144 & 1.000000 & 0.201677 & 0.312334 \\
Claude Sonnet 4.6 & high & 0.3432 & 1501.60 & 0.600402 & 0.150794 & 0.034454* & 0.035554* \\
Claude Sonnet 4.6 & low & 0.4597 & 1025.11 & 0.332112 & 0.690476 & 0.057008 & 0.019168* \\
Gemini 3 Flash & high & 0.3419 & 1174.01 & 0.117525 & 0.666430 & 0.828812 & 0.179447 \\
Gemini 3 Flash & low & 0.2501 & 784.67 & 0.599286 & 0.595883 & 0.372748 & 0.292211 \\
Gemini 3.1 Pro & high & 0.4991 & 1229.54 & 0.420635 & 0.222222 & 0.007937* & 0.007058* \\
Gemini 3.1 Pro & low & 0.1004 & 211.67 & 1.000000 & 1.000000 & 1.000000 & 0.996306 \\
Mistral Large & default & 0.1635 & 705.14 & 0.294802 & 1.000000 & 0.666430 & 0.363966 \\
GPT-5.2 & low & 0.1128 & 200.37 & 0.423711 & 0.007495* & 1.000000 & 0.000849* \\
GPT-5.2 & xhigh & 0.6501 & 1244.53 & 0.083265 & 0.031141* & 0.151494 & 0.158790 \\
GPT-5.4 & low & 0.3730 & 711.01 & 0.205903 & 0.547619 & 0.142457 & 0.393770 \\
GPT-5.4 & xhigh & 0.8748 & 899.42 & 1.000000 & 0.007495* & 0.423711 & 0.004349* \\
GPT-5.4 mini & low & 0.0027 & 88.51 & 0.423711 & 1.000000 & 1.000000 & 0.391625 \\
GPT-5.4 mini & xhigh & 0.4252 & 1262.49 & 0.528359 & 0.841270 & 0.753298 & 0.596476 \\
\bottomrule
\end{tabular*}
\end{table}
\end{landscape}

These findings reveal condition sensitivity, but not in a way that would justify
treating environmental shifts as the primary source of risk. For most
variants, the more substantial problem is that repeated runs under the
same condition still yield inconsistent quality.

\section{Discussion}\label{discussion}

\subsection{Main findings}\label{main-findings}

Taken together, the present evidence does not support a broad claim that
current LLM agents are ready to serve as reliable autonomous EDA experts
in this task setting. Most evaluated model-variants either perform too
weakly, vary too strongly within the same condition, or show sensitivity
to at least one environmental intervention. The results also indicate
that reasoning effort changes model behavior. In several
families, higher-effort variants perform much better than their
lower-effort counterparts, with the clearest examples being
GPT-5.4 and GPT-5.2. However, that
effect is not uniform across all families or variants, which suggests
that additional reasoning budget changes performance substantially
without reliably producing stable analytical competence.

At the same time, the study shows that forms of instability are not
equally important. Cross-experiment sensitivity is informative because
it reveals where model behavior changes when the analytical environment
changes. However, it is not the dominant source of risk in this
benchmark. The larger issue for most configurations is within-experiment
unpredictability, captured by high \metric{CV} and translated into a
weak stability discount. Many variants appear not to apply a robust
exploratory strategy consistently. Instead, repeated runs under the same
condition often diverge between productive and unproductive lines of
reasoning, which suggests unstable strategy discovery rather than
reliable problem understanding. This pattern remains visible even for
configurations given substantial reasoning effort and execution time.

Across model-variants, the average \businessutility{} is
\texttt{0.1511}, whereas the average \metric{ms} is \texttt{0.3562},
which means the risk-adjusted summary is only about \texttt{42\%} as
large as the mean-score summary. Mean performance therefore overstates
practical readiness when instability is high.

\subsection{Relation to EDA Theory}\label{relation-to-eda-theory}

The benchmark findings also matter specifically for EDA rather than only
for generic task completion. In classical EDA, the analyst is not simply
retrieving a known answer but forming and refining plausible hypotheses
from partial structure in the data
(\citeproc{ref-de_mast_exploratory_2007}{De Mast and Trip 2007};
\citeproc{ref-good_philosophy_1983}{Good 1983};
\citeproc{ref-tukey_exploratory_1977}{Tukey 1977}). That process depends
on consistency. An analyst who describes the same data differently on
each pass, notices different salient features in each repeated attempt,
or interprets them inconsistently would not be regarded as dependable.
The same standard should apply to LLM-based EDA agents.

This point helps explain why within-condition variance is so damaging in
the present benchmark. The task is not ambiguous, as evidenced by
multiple perfect scores achieved by some model-variants. The simulation
was intentionally configured to contain repeated signal traces that make
near-optimal performance possible. The problem is therefore not that the
data are too noisy to support correct reasoning, but that most models do
not reliably discover and execute a workable exploratory strategy.

\subsection{Relation to Prior Benchmark
Literature}\label{relation-to-prior-benchmark-literature}

The findings are also consistent with a broader trend in the recent
benchmark literature. Benchmarks such as \texttt{DSBench},
\texttt{IDA-Bench}, \texttt{DA-Code}, \texttt{DAComp}, and
\texttt{Spider2-V} all suggest that average task success overstates
practical readiness when tasks become realistic, multi-step, or
tool-grounded (\citeproc{ref-cao_spider2-v_2024}{Cao et al. 2024};
\citeproc{ref-huang_da-code_2024}{Y. Huang et al. 2024};
\citeproc{ref-jing_dsbench_2025}{Jing et al. 2025};
\citeproc{ref-lei_dacomp_2025}{Lei et al. 2025};
\citeproc{ref-li_ida-bench_2025}{Li et al. 2025}). \texttt{InsightEval}
and \texttt{FDABench} similarly show that evaluating analytical agents
requires richer metrics than one scalar accuracy score because insight
quality, workflow quality, and heterogeneous evidence handling all
matter (\citeproc{ref-wang_fdabench_2025}{Wang et al. 2025};
\citeproc{ref-zhu_insighteval_2025}{Zhu et al. 2025}).

The present study extends that literature in two ways. First, it narrows
the focus to a process-oriented EDA problem rather than a broader
data-agent task bundle. Second, it shows that a benchmark can look more
favorable than it really is if repeatability is ignored. The high
variance observed here suggests that the practical gap between
occasional analytical success and dependable analytical use remains
substantial even when the underlying task structure is strong enough to
permit near-perfect answers.

\subsection{Implications for
Deployment}\label{implications-for-deployment}

From a deployment perspective, the results support a cautious
interpretation of current LLM-based analytics. Organizations should
realistically assess whether a model performs well enough, often enough, and
predictably enough to reduce rather than increase verification burden.
An equally important question is what level of
\businessutility{} would be sufficient for managers in real
organizations to accept EDA agents as viable substitutes for human
analysts in at least some exploratory tasks. The present benchmark helps
structure that question, but it does not yet establish a deployment
threshold.

\section{Conclusion}\label{conclusion}

This study leads to a cautious but substantively important conclusion
about LLM-based EDA agents. In the present benchmark, the central
obstacle to practical adoption is not merely that many models fail to
reach a sufficiently high mean score. The deeper problem is that average
performance is often coupled with high within-condition dispersion and,
for some variants, selective sensitivity to changes in the analytical
environment. In operational terms, that means a model may appear
competent on average while still remaining unreliable in repeated use.
For exploratory analysis, this is a critical weakness, because EDA is
valuable only when the analyst can repeatedly recover meaningful
structure from the same underlying evidence rather than alternate
unpredictably between productive and unproductive reasoning paths.

The paper contributes to model-evaluation methodology. Rather than
treating \metric{ms} as a sufficient indicator of readiness, it proposes
a reporting framework in which \metric{ms}, \metric{CV},
\businessutility{}, and condition sensitivity play distinct but
complementary roles. \metric{ms} captures average analytical quality,
\metric{CV} captures unpredictability, \businessutility{}
translates quality and repeatability into a single risk-adjusted
summary, and the non-parametric cross-condition tests provide an
auxiliary view of environmental sensitivity. This framework matters
because it shifts evaluation away from a narrow question of whether a
model can produce a correct answer at all and toward a more
decision-relevant question, whether it can do so often enough and
consistently enough to justify trust in realistic use. In that sense,
the study is not only an empirical benchmark of current models, but also
a methodological argument for evaluating analytical agents in a way that
is closer to organizational decision criteria.

The role of \businessutility{} is especially important in that
contribution. The metric is proposed as a utility-theoretic bridge
between raw benchmark outcomes and practical judgments about usefulness.
Its prospect-theory-inspired discount does not claim to model managerial
behavior directly, but it encodes a plausible operational preference -
instability should reduce perceived value because unstable systems
impose verification costs and make failure harder to anticipate. This
point connects the benchmark to a broader question of AI adoption in
organizations (\citeproc{ref-labedzki_human-ai_2025}{Łabędzki et al.
2025}). Managers do not deploy analytical systems on the basis of
average capability alone. They deploy systems when expected gains are
high enough, and predictable enough, to support trust, accountability,
and acceptable oversight burden. From that perspective, the benchmark
suggests that trust in AI EDA agents should be understood not as a vague
attitude, but as a function of demonstrated analytical quality,
repeatability, and robustness under changing conditions.

Taken together, the findings suggest that current frontier models have
not yet closed the gap between occasional analytical success and
dependable analytical use in this task setting. The strongest
configuration, GPT-5.4 xhigh, shows that
high-performing and relatively stable EDA-like behavior is possible.
However, the broader pattern remains one of operational fragility, with
most variants still too unstable to support confident substitution for
human analysts. The most important implication is therefore twofold.
Substantively, organizations should be cautious about interpreting
isolated strong runs or benchmark averages as evidence of deployment
readiness. Methodologically, future evaluations of analytical agents
should treat trustworthiness, usefulness, and adoption potential as
multidimensional constructs that require joint measurement of efficacy,
repeatability, and environmental sensitivity rather than reliance on any
single scalar score.

\section{Limitations}\label{limitations}

The study has several important limitations. First, each model-variant
was evaluated on only five trajectories per experiment, which limits the
power of non-parametric significance tests and increases uncertainty
around variance estimates. Second, the benchmark covers one business
problem and one simulation family rather than the full spectrum of EDA
work. Third, the final usefulness of \businessutility{} depends
on the chosen prospect-theory-inspired parametrization and on the
decision to embed the resulting loss term inside an exponential
discount. Fourth, the accepted convention \texttt{CV\ =\ 0} for
zero-score groups keeps the computation defined but makes the derived
stability discount less informative for collapsed states.

\section*{Acknowledgements}\label{acknowledgements}

The authors acknowledge the additional contributors to this research:
Cezary Angielczyk, Bryan Chen, Mikołaj Drojma, Jakub Filarecki, Maciej
Kaczkowski, Jan Klamka, Wiktor Kopczyński, Paweł Marcinkowski, Dorota
Mockiewicz, Jędrzej Nogacki, Alicja Nowakowska, Mateusz Siwak, and
Marcin Umiński.

\section*{Use of artificial intelligence}\label{use-of-artificial-intelligence}

The authors used artificial intelligence tools in preparing this manuscript.
They were employed for text correction and proofreading, for support in
data analysis, and for assistance with \LaTeX{} formatting. The authors
reviewed all outputs and remain responsible for the content.

\section*{Data Availability}\label{data-availability}

The benchmark dataset and supporting project materials are publicly
available in the project repository,
\href{https://github.com/deepsense-ai/agent-based-simulation-benchmark}{deepsense-ai/agent-based-simulation-benchmark}.

\section*{References}\label{references}

\protect\phantomsection\label{refs}

\end{document}